# CW RF SYSTEM OF THE PROJECT-X ACCELERATOR FRONT END*

T. Khabiboulline, S. Barbanotti, I Gonin, N. Solyak, I. Terechkine and V. Yakovlev,
FNAL, Batavia, IL 60510, USA.

*Abstract*

Front end of a CW linac of the Project X contains an H⁻ source, an RFQ, a medium energy transport line with the beam chopper, and a SC low-beta linac that accelerates H⁻ from 2.5 MeV to 160 MeV. SC Single Spoke Resonators (SSR) will be used in the linac, because Fermilab already successfully developed and tested a SSR for beta=0.21. Two manufactured cavities achieve 2.5 times more than design accelerating gradients. One of these cavities completely dressed, e.g. welded to helium vessel with integrated slow and fast tuners, and tested in CW regime. Successful tests of beta=0.21 SSR give us a confidence to use this type of cavity for low beta (0.117) and for high-beta (0.4) as well. Both types of these cavities are under development. In present report the basic constrains, parameters, electromagnetic and mechanical design for all the three SSR cavities, and first test results of beta=0.21 SSR are presented.

## INTRODUCTION

Project X is a high intensity proton accelerator conceived to support a world-leading program in neutrino and flavour physics over the next two decades at Fermilab [1]. A CW linac provides several important advantages to the rare processes experimental program while preserving the beam characteristics for the long baseline neutrino program. The beam quality and the duty factor of a CW linac are significantly better than that for slow extracted beams. The beam originates from a 1-10 mA DC H⁻ source. The beam is then bunched and accelerated by a CW normal-conducting RFQ to 2.5 MeV, and then bunches are formatted by a chopper following a pre-programmed timeline. From 2.5 MeV to 3 GeV the H⁻ bunches are accelerated by a CW super-conductive linac. The beam average current in the linac is 1 mA. The linac energy of 3 GeV is sufficient to meeting the requirements of both the Mu2e and a Kaon programs.

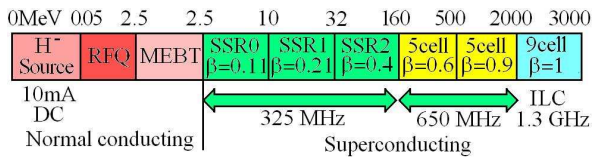

Figure 1: Layout of the Project X Linac.

## NORMAL CONDUCTING PART

The H⁻ ion source provides 10mA DC current. The transverse emittance is required to be less than 0.25 πmm-mrad (rms, norm) and beam halo must be controlled to prevent unacceptable beam losses at high energies. The bunch repetition rate within a train is 325 MHz, however not all of the RF buckets would contain bunches. There are two principal time structures needed for the linac operation: (1) pulsed, 5ms at 10 Hz and (2) CW with a bunch structure variation at a microseconds level. The pulsed time structure is required to provide 4.3-ms long trains of bunches at 10 Hz rate for the injection into the Rapid Cycling Synchrotron (RCS) or pulsed 1.3 GHz CS linac, where the beam is accelerated up to 8 GeV. If RCS is used, the bunch structure must incorporate the RCS RF bucket frequency (50.33 MHz) structure to facilitate pseudo bunch-to-bucket transfer and also the RCS revolution frequency (0.513 MHz) structure to provide a 200-ns extraction gap in the RCS ring. The CW bunch structure is determined by rare decay experiments. In summary, the beam bunches can occupy arbitrary RF buckets as long as the average current does not exceed 1 mA. It is likely that such a time structure will be provided by two choppers, the first one (a pre-chopper) immediately after the ion source and the second at MEBT section after the beam acceleration in the RFQ.

The CW RFQ section provides bunching of the beam at 325 MHz and acceleration up to 2.5 MeV. There is no SC RFQ in operation for protons/H⁻ (80 MHz, 350 keV SRFQ for heavy ions operates for PIAVE linac). On the another hand, there is a number of CW RFQs operating at room temperature for beam currents of up to 100 mA and energies of up to ~7 MeV. CW 350 MHz RFQ is developed for the output energy of 3 MeV and beam current of 20 mA for KOMAC/KTF. The wall losses are 350 kW, and the length is 3.3 m. The accelerator is built and is under commissioning. Design and experience of this accelerator may be used as well. The RF source for a CW RFQ at ~1 MW power is available at 350 MHz; it is possible to redesign it for 325 MHz.

## SUPERCONDUCTING LOW BETA PART

Initial concept of the Project X [2] was based on the pulse linac. For the CW concept the major modification in the low energy (325-MHz) part compared to the pulsed concept includes replacing 16 room temperature cross-bar cavities with the SC spoke cavities. The room temperature cavities were designed for pulsed linac with duty factor of up to 1% and they cannot be used in CW regime. In Fermilab we successfully develop of 325 MHz Superconducting Spoke Resonators SSR1 for beta 0.21. One dressed cavity reached 27 MV/m accelerating gradient at 4.5K and second undressed cavity quenched at 33 MV/m at 2K. Based on this experience SSR0 cavity was designed for beta 0.11 to replace normal conducting cavities. The RF design of the beta 0.4 SSR2 cavities is already done.

---

*Work supported by US DoE
#khabibul@fnal.gov

Work supported by Fermi Research Alliance, LLC under Contract No. DE-AC02-07CH11359 with the US Dept. of Energy

Table 1: Main parameters of the SSR part of the CW linac

| Section | Energy range MeV | β | Power/cavity, kW ($I_{av}$=1 mA) |
|---|---|---|---|
| SSR0 | 2.5-10 | 0.073-0.146 | 0.5 |
| SSR1 | 10-32 | 0.146-0.261 | 1.3 |
| SSR2 | 32-160 | 0.261-0.5 | 4.1 |

*SSR0 cavity design.*

CW SSR0 cavities will be used for acceleration of the beam energy from 2.5 MeV to 10 MeV. The RF parameters of the SSR0 cavity for β =0.073-0.146 were optimized. Optimization parameters were distribution of electric and magnetic fields in optimized cavity, cavity stiffness and length, etc. The results of optimization are shown in Table 2.

Table 2: SSR0 cavity parameters.

| Operating frequency | 325 MHz |
|---|---|
| Optimal Beta, $β_{opt}$ | 0.114 |
| Qload | $6.5 \cdot 10^6$ |
| $E_{peak}/E_{acc}$ [1] | 5.63 |
| $B_{peak}/E_{acc}$ [1] | 6.92 mT/(MV/m) |
| G | 50 Ω |
| R/Q | 108 Ω |
| Cavity effective length, $D_{eff} = 2*βλ/2$ | 105mm |

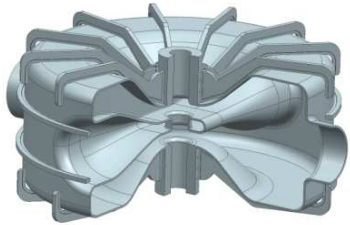

Figure 2: Concept of the SSR0 mechanical design

Due to longitudinal restrictions, SSR0 cavity RF design doesn't take the advantages of re-entrant shape similar to SSR1. The magnetic field enhancement factor is about 16% higher. The HOM analysis, power coupler and mechanical design of SSR0 are underway. Note that for CW operation with low beam loading, more important is not Lorentz factor but microphonics: the peak amplitude of microphonics should not be greater than 30 Hz. Thus, frequency tolerance versus the pressure fluctuation df/dP (f – frequency, P – He pressure) should not exceed ~50-70 Hz/Torr in order to sustain the maximal He pressure fluctuation of 0.4 Torr. Same time the cavity spring constant should allow the cavity tuning. All these factors need to be taken into account during the development of the cavity mechanical design. Figure 2 shows the present design for the SSR0 cavity. The proposed design includes 12 radial ribs in the end wall area and 2 circumferential ribs on the outer shell. The rib design satisfies the requirements to minimize the sensitivity to pressure fluctuations during operation and to safely operate the cavity in a cryostat. Different rib configurations are being studied, both on the mechanical and RF point of view. Further studies will lead to the final design, but the present proposal is already satisfactory.

Concept of a design lattice structure that includes the SSR0 cavity and focusing solenoid is shown in Figure 3.

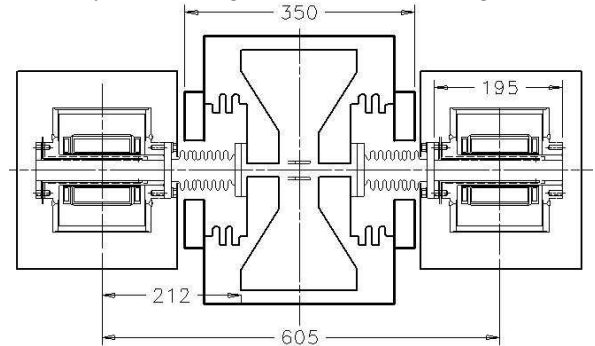

Figure 3: Design Lattice structure including SSR0 cavity and focusing solenoids.

*SSR1 cavity development status*

The second family of spoke resonators SSR1 are the same as in the ICD-1 8 GeV linac and will accelerate the beam from 10 MeV to 32 MeV. Parameters of the cavities are shown in Table 3.

Table 3: SSR1 cavity parameters.

| Operating frequency | 325 MHz |
|---|---|
| Optimal Beta, $β_{opt}$ | 0.215 |
| Qload | $6.5 \cdot 10^6$ |
| $E_{peak}/E_{acc}$ [1] | 3.84 |
| $B_{peak}/E_{acc}$ [1] | 5.81 mT/(MV/m) |
| G | 84 Ω |
| R/Q | 242 Ω |
| Cavity effective length, $D_{eff} = 2*βλ/2$ | 198.5mm |

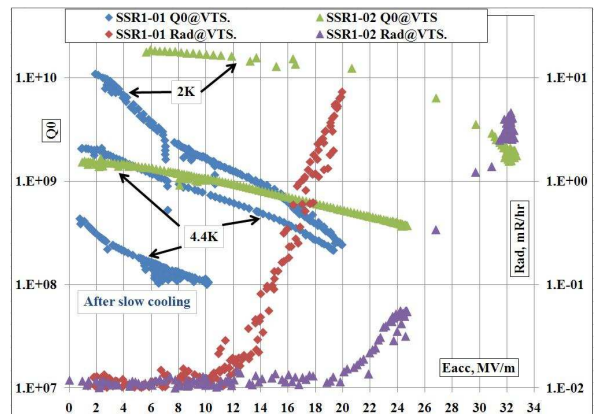

Figure 4: $Q_0$ vs. acceleration gradient $E_{acc}$ and x-rays from the cold tests of SSR1-01 and SSR1-02 in VTS.

Single spoke 325 MHz cavity SSR1 having $β_G$ =0.21 was designed and built for the HINS project. Two cavities were already manufactured in two different companies, ZANON from Italy and ROARK from USA. After bulk BCP and High Pressure Water Rinsing at ANL both cavities were RF tested in Fermilab VTS at 4.4K and 2K. Results of the cavity tests are shown in Figure 4. Cavity SSR1-01 then was baked at 600C during 10 hours at JLAB, flash BCP and HPWR at ANL, tuned to operate at 4.6K and welded to Helium vessel. After installation of high Qext power coupler and two tuners cavity SSR1-01 was installed in HINS Cavity Test Facility Cryostat (CTFC) and tested at 4.6K, see Figure 5.

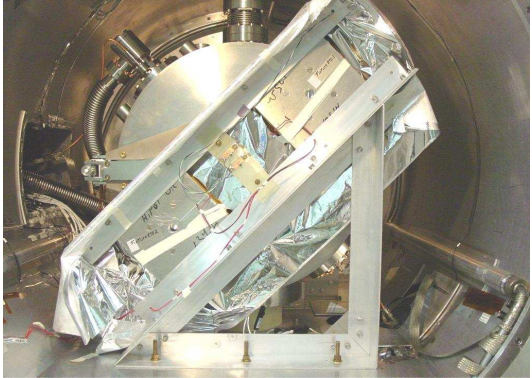

Figure 5: SSR1-01 cavity in CTFC.

In CTFC cavity accelerating gradient reached 27 MV/m limited by quench. Improvement from 19 MV/m reached in VTS is the result of baking and additional BCP, see Figure 6. In the Project-X linac focusing solenoids are located close to SSR cavity. Four coils installed inside of cryostat near the cavity. They allow us to apply external magnetic field 2Gs on cavity surface per 1A of current of solenoids. Study of Cavity Q0 degradation under external DC magnetic field was performed. Under 10 Gs of external magnetic field after 5000 quenches no change of Q0 was observed [3].

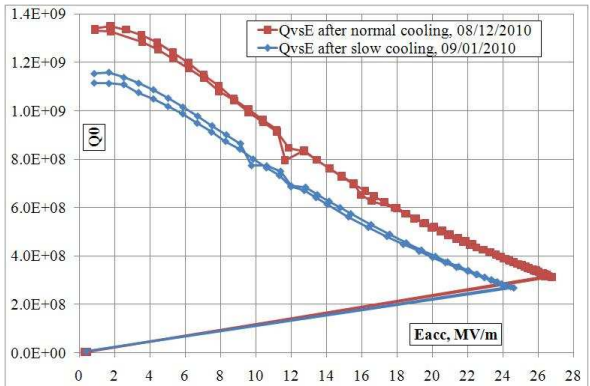

Figure 6: $Q_0$ vs. acceleration gradient $E_{acc}$ of the SSR1-01 cavity in CTFC after fast and slow cooling down.

Preliminary lattice design is shown in Figure 7. It includes fully dressed SSR1 cavity with two tuners between focusing solenoids. Distance between centres of solenoids define focusing period of 800 mm.

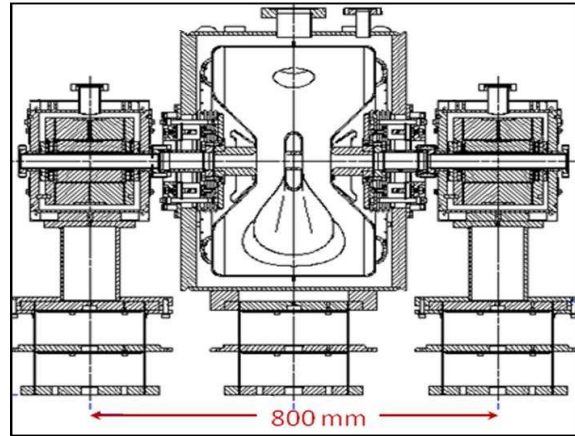

Figure 7: One cell of a lattice design for SSR1 section.

*SSR2 cavity design*

The third family of spoke resonators SSR will be used for acceleration of the beam energy from 32 MeV to 160 MeV. The RF parameters of the SSR0 cavity for β =0.261-0.5 was optimized for 2 different beam pipe diameter 30 and 40 mm. The results of optimization are shown in Table 4.

Table 4: SSR2 cavity parameters.

| Operating frequency | 325 MHz | 325 MHz |
|---|---|---|
| Beam pipe diameter | 30mm | 40mm |
| Optimal Beta, $β_{opt}$ | 0.414 | 0.419 |
| Qload | $1\,10^7$ | $1\,10^7$ |
| $E_{peak}/E_{acc}$ [1] | 3.42 | 3.67 |
| $B_{peak}/E_{acc}$ [1] | 6.81 mT/(MV/m) | 6.93 |
| G | 112 Ω | 109 |
| R/Q | 304 Ω | 292 |
| Cavity effective length, $D_{eff}=2*βλ/2$ | 381.9mm | 386.5 |

## FUTURE PLANS

Power coupler of the cavity SSR1-01 will be replaced by high power coupler with $Q_{ext}$=2.5e6. Then cavity will be installed in HINS HTS and tested in pulsed regime. 40 kW power is available for pulsed test.

Cavity SSR-02 will be prepared for welding to Helium vessel. Additional 12 SSR1 cavities are under production.

Mechanical design of the SSR0 cavity nearly finished and first cavity production will be started next year.

Test cryostat for 4 cavities and 5 solenoids design work started.